# MHD Precursor to disruption in Iran Tokamak 1

Alireza Hojabri[1,2], Mahmmod Ghoranneviss[2] and Fatemeh Hajakbari[1,2]

*Physics group, Islamic Azad University of Karaj 31485-313, Iran.*
*Plasma Physics Research Center, Azad University, Tehran 14835-197, Iran*

The purpose of this paper is to further investigate the major disruptions occurring in low-q(a) discharges in Iran Tokamak 1, and to compare the theoretical and experimental results for the rate of island growth. The study of precursor phase of disruption can be predicted and avoided using suitable control systems. In this paper are described the stability analysis and the observed growth rates indicate that the rotating modes are tearing modes.

**Introduction**

Tearing mode is known of importance in its relation to the degradation of the confinement and the disruption in Tokamak [1,2]. Tearing mode is a resistive instability. It is driven by the free energy in the poloidal magnetic field. They are resonant instabilities, localized around flux surfaces and change the topology of the magnetic flux distribution through the formation of magnetic islands. There are presently distinct MHD models of Tokamak disruption. The first model was that the interaction of the single helicity m=2/n=1 magnetic island with the limiter or cold gas region, ruins the energy confinement [3]. In second scenario, disruptions believed to be triggered by nonlinear coupling of the modes of different helicities. In cases where both m=2/n=1 and m=3/n=2 modes are unstable and excite a broad spectrum of waves which destroy magnetic surfaces and therefore strongly enhance cross field thermal collapse in the region of the stochastic magnetic field [4]. In the third model of disruptions was proposed namely a nonlinear interaction between the m=1/n=1 and

m=2/n=1 modes taking place through the intermediary of the current profile [5,6].It is important to understand in detail the various mechanism that can lead to disruption, not only important in Tokamak confinement ,but also in suppressing of such disruption in the tokamak [7].

**Experimental Results and Discussion**

A typical shot which include major disruption is illustrated in fig.1. In this shot, the relevant major disruption (q(a)=2.5) occurs at 13.5ms. Their main characteristics are : negative spikes in the loop voltage signal (fig.1.b), rapid loss of confined energy (fig.1.e) and sudden expansion of the minor radius with a diminution of the major radius(fig.1.c). It is furthermore typically verified that there is an intense MHD activity with a frequency decrease and fast growth in amplitude (fig.1.e), associated with the occurrence of an major disruption. The major disruptions in IR-T1 discharges with 2<q(a)<3 are usually characterized by a noticeable growing precursor activity, several negative spikes in the loop voltage signal and a displacement of the plasma column towards the inner part of the torus and interact with the limiter, which is evident by pulsed increases in the emission of the $H_\alpha$ signal ($\lambda$=6563A,3-2). As the plasma is shifted inwards, the inner limiter scrapes off the plasma, reducing the radius of the current channel and thus enhancing the resistance. This means a steadily increases heat loading on the inside of the limiter,which could be the cause of the increases in $H_\alpha$ radiation. The increase of $H_\alpha$ radiation is clearly associated with a thermal collapse of the plasma boundary, of course the stochastic regions that can lead to the thermal collapse.

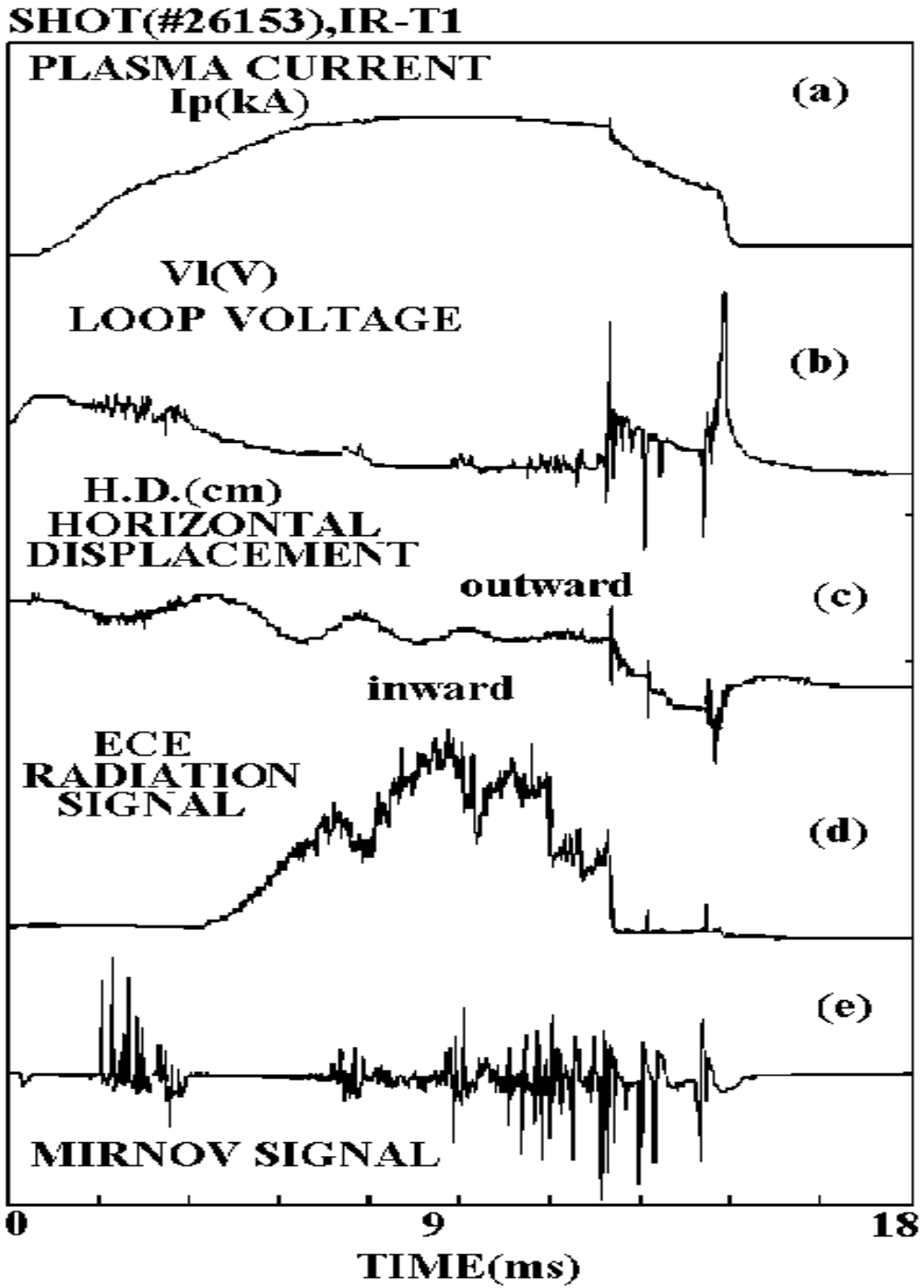

Figure 1. Time behavior of principal parameters (Shot No. 153) in the major disruption event on the IR-T1 tokamak; (a) plasma current (kA), (b) loop voltage (volt), (c) horizontal displacement signal (cm), (d) ECE signal (arb. units), (e) Mirnov coil signal (arb. units). The disruptive instability is appeared close to 13.5ms.

**Conclusion**

In this work we describe detailed measurements made of disrupting hydrogen plasmas in ohmic heating. The major disruption studied here consist of three phase : 1) Thermal instability, 2) Growth of the m=2 magnetic island, 3) Trigger mechanism. The other conclusion from this study is that the magnitude of tokamak m=2 poloidal fluctuations (Mirnov oscillations) dependence on the value of the safety factor at the limiter.